\documentclass[final]{IEEEtran}
\usepackage{cite}
\usepackage{graphicx}
\usepackage{picinpar}
\usepackage{stfloats}
\usepackage{setspace}
\usepackage[all]{xy}
\usepackage{float}
\usepackage{array}
\usepackage{mdwmath}
\usepackage{mdwtab}
\usepackage{eqparbox}
\usepackage{graphicx}
\usepackage{picinpar}
\usepackage[cmex10]{amsmath}
\usepackage{amsmath,amsfonts,amssymb}
\usepackage{algorithmic}
\usepackage[linesnumbered,ruled,lined,boxed]{algorithm2e}
\usepackage{subfigure}
\usepackage{threeparttable}
\usepackage{caption}
\usepackage{booktabs}
\usepackage{multirow}
\usepackage{etoolbox}
\usepackage{makecell}
\usepackage{amssymb}
\usepackage{mathrsfs}
\usepackage{color}
\usepackage{diagbox}
\definecolor{purple}{RGB}{160,32,240}
\definecolor{darkred}{RGB}{255,0,255}
\usepackage{stfloats}
\usepackage{bm}

\definecolor{purple}{RGB}{160,32,240}
\definecolor{darkred}{RGB}{255,0,255}

\newcommand{\e}{\begin{equation}}
\newcommand{\ee}{\end{equation}}
\newcommand{\eqn}{\begin{eqnarray}}
\newcommand{\eeqn}{\end{eqnarray}}
\begin{document}
{\title{\Large{Joint Activity Detection and Channel Estimation for Massive IoT Access Based on Millimeter-Wave/Terahertz Multi-Panel Massive MIMO}}}
\author{
Hanlin Xiu, Zhen Gao, Anwen Liao, Yikun Mei, Dezhi Zheng, Shufeng Tan,  Marco Di Renzo,~\IEEEmembership{Fellow,~IEEE}, and Lajos Hanzo,~\IEEEmembership{Fellow,~IEEE}%
\vspace{-8mm}

}
\maketitle

\begin{abstract}
 The multi-panel array, as a state-of-the-art antenna-in-package technology, is very suitable for millimeter-wave (mmWave)/terahertz (THz) systems, due to its low-cost deployment and scalable configuration.
 But in the context of non-uniform array structures it leads to intractable signal processing.
 Based on such an array structure at the base station, this paper investigates a joint active user detection (AUD) and channel estimation (CE) scheme based on compressive sensing (CS) for application to the massive Internet of Things (IoT).
 Specifically, by exploiting the structured sparsity of mmWave/THz massive IoT access channels, we firstly formulate the multi-panel massive multiple-input multiple-output (mMIMO)-based joint AUD and CE problem as a multiple measurement vector (MMV)-CS problem.
 Then, we harness the expectation maximization (EM) algorithm to learn the prior parameters (i.e., the noise variance and the sparsity ratio) and an orthogonal approximate message passing (OAMP)-EM-MMV algorithm is developed to solve this problem.
 Our simulation results verify the improved AUD and CE performance of the proposed scheme compared to conventional CS-based algorithms.
\end{abstract}

\IEEEpeerreviewmaketitle

\vspace{-1mm}
\section{Introduction}\label{S1}

 Multi-panel massive multiple-input multiple-output (mMIMO) is a viable array configuration to realize the future millimeter-wave (mmWave)/terahertz (THz) communications \cite{WZhang_CM18,WZhang_Tcom20}.
 Specifically, the antenna elements are integrated into a uniform planar array (UPA) to create a panel, and multiple panels are juxtaposed to form the multi-panel mMIMO array shown in Fig.~\ref{FIG1}.
 As a partially-connected hybrid MIMO architecture relying on a modest number of RF chains, multi-panel mMIMO schemes exhibit high energy efficiency \cite{WZhang_Tcom20}.
 Moreover, compared to conventional mMIMO arrays having half-wavelength antenna spacing, multi-panel arrays have advantages of low-cost deployment and flexible configurations \cite{WZhang_Tcom20}.
However, the resultant non-uniformly spaced arrays pose challenging on signal processing\cite{WDM_TVT20}.

 In addition, the next-generation communications are expected to support the high-throughput uplink transmission, including the applications of Internet of Things (IoT), Internet of Vehicles (IoV), and meta-universe, where efficient massive IoT access protocols are a prerequisite \cite{YsL_TVT20_short,YsLHanzo_TVT20_long,ZZY_JSAC19}.
 Sophisticated techniques have been proposed in the literature \cite{Shim_SPAWC17,VLau_ICC15,SXD_IoTJ19,SXD_TSP20,Shim_Tcom19,ZCYW_TSP18,KML_TSP20} for the joint active user detection (AUD) and channel estimation (CE) in support of massive IoT access.
 In \cite{VLau_ICC15}, by exploiting both the active user sparsity and the joint sparsity structures observed at multiple receive antennas, a modified Bayesian compressive sensing (CS) algorithm was proposed for joint AUD and CE.
 The authors of \cite{Shim_SPAWC17} proposed an orthogonal matching pursuit (OMP)-based joint AUD and time-domain CE technique for grant-free massive IoT access.
 Similar to other greedy algorithms, this detector fails to effectively harness any {\emph{a priori}} information, and the associated high-dimensional matrix inversion imposes excessive complexity.
 To reduce the complexity, an approximate message passing (AMP) algorithm based joint AUD and CE scheme was developed in \cite{ZCYW_TSP18}, but this AMP design requires the prior distributions of wireless channels and the noise variance to be known, which are hard to acquire in practice.
 In \cite{Shim_Tcom19}, by exploiting both the active user sparsity and the joint sparsity observed at the multiple receive antennas, an efficient low-complexity expectation propagation-based algorithm was proposed under the Bayesian framework for joint AUD and CE.
 {\color{black}In \cite{Shim_IoTJ21}, the authors proposed a deep learning based AUD and CE in the grant-free non-orthogonal multiple access (NOMA) systems, where deep learning figured out the direct mapping between the received NOMA signal and the indices of active devices and associated channels using the long short-term memory.}
 However, the schemes in \cite{Shim_SPAWC17,VLau_ICC15,Shim_Tcom19,ZCYW_TSP18,YsL_TVT20_short,YsLHanzo_TVT20_long,Shim_IoTJ21} have not considered mMIMO systems.
 As a further advance, the authors of \cite{SXD_IoTJ19} designed an mMIMO-based three-phase transmission protocol, which consist of joint AUD and CE conceived for uplink and downlink data transmission in massive cellular IoT access.
 To solve the joint AUD and CE problem in grant-free random access over a given coherence interval, the authors of \cite{SXD_TSP20} proposed a logarithmic smoothening method for handling a non-smooth objective function.
 Based on the structured sparsity of the channel matrix, a generalized multiple measurement vector (GMMV)-AMP algorithm was proposed for the uplink of broadband massive IoT access systems \cite{KML_TSP20}.
 However, the fully-digital mMIMO considered in \cite{SXD_IoTJ19,SXD_TSP20,KML_TSP20} suffer from prohibitively high hardware cost and power consumption.
 We provide a brief summary of the related literature in Table I.
 Furthermore, when the sensing matrices are ill-conditioned, the mean square error (MSE) performance and the convergence speed of the orthogonal AMP (OAMP) algorithm proposed in \cite{MaJ_Access17} outperforms the existing AMP algorithms.
{\color{black} However, the conventional OAMP algorithm is restricted to the single measurement vector (SMV) CS problem. Moreover, the OAMP algorithm requires the {\it {a priori}} distribution to be known, whose parameters are difficult to obtain in the realistic communication systems.}

\begin{table}[!t]
\tiny
\centering
\captionsetup{font = {normalsize, color = {black}}, labelsep = period} 
\caption*{Table I: A brief comparison of the related literature}
\begin{tabular}{|c|c|c|c|c|c|}
\Xhline{0.8pt}
\multicolumn{2}{|c|}{\diagbox{Contents}{Literature}} &\cite{WZhang_Tcom20,WDM_TVT20}  &\cite{Shim_SPAWC17,VLau_ICC15,Shim_Tcom19,ZCYW_TSP18,Shim_IoTJ21}&\cite{SXD_IoTJ19,SXD_TSP20,KML_TSP20}&Proposed \\%
\Xhline{0.8pt}
\multirow{3}*{BS}
&1/2/4 Antennas & &  \checkmark &  & \\
\cline{2-6}
&Fully-digital mMIMO & & &\checkmark  &  \\
\cline{2-6}
&Multi-panel mMIMO  &\checkmark(Linear Array) & &   &\checkmark(Planar Array) \\
\Xhline{0.8pt}
\cline{2-6}
\multirow{2}*{Processing at BS}
&CE    &\checkmark &\checkmark &\checkmark  & \checkmark  \\
\cline{2-6}
&AUD   &  &\checkmark &\checkmark & \checkmark  \\
\Xhline{0.8pt}
\end{tabular}
\vspace{-1mm}
\end{table}

 {\color{black} In this paper, we study the multi-panel mMIMO operating at mmWave/THz frequency for high-throughput  massive IoT access.
 Specifically, a CS-based joint AUD and CE scheme is proposed in support of the high-efficient uplink access, where the multi-panel MIMO array at the BS adopts a partially-connected hybrid architecture.
 We introduce the mmWave/THz multi-panel mMIMO channel model for the first time.
 By exploiting the structured sparsity of massive IoT access channels, the joint AUD and CE problem can be formulated as a multiple measurements vector (MMV) problem under the CS framework.
 To solve this MMV-CS problem in the massive IoT access based on the multi-panel mMIMO system, we develop an OAMP-expectation maximization (EM)-MMV algorithm, where the EM algorithm can adaptively learn some unknown parameters, i.e., the noise variance and the sparsity ratio.
 Moreover, the sensing matrix of the multi-panel system can be easily designed to be a partially unitary matrix, so that the computational complexity of the proposed OAMP-EM-MMV algorithm can be reduced and the signal processing challenges of the associated non-uniform array can be mitigated.
 Finally, our simulation results verify that the OAMP-EM-MMV algorithm proposed for joint AUD and CE has a better performance than conventional CS-based algorithms.
 }

\textit{Notations}:
 Boldface lower and upper-case symbols denote column vectors and matrices, respectively.
 The superscripts $ (\cdot)^{\rm T}$, $(\cdot)^{\rm H} $, and $(\cdot)^{-1} $ denote the transpose, conjugate transpose, and matrix inversion operators, respectively;
 ${\| {\bf{a}} \|_2}$ and ${\| {\bf{A}} \|_F}$ are the ${\ell_2}$-norm of ${\bf{a}}$ and the Frobenius norm of ${\bf{A}}$, respectively;
 $\otimes$ denotes the Kronecker product operation;
 ${\bf{0}}_{N}$ and ${\bf{I}}_{N}$ represent the vector of size $N$ with all the elements being 0 and the $N\! \times\! N$ identity matrix, respectively;
 ${\rm vec}[{\bf{A}}]$ stacks the columns of ${\bf{A}}$ on top of each other;
 ${\rm tr}({\bf{A}})$ is the trace of ${\bf{A}}$ that calculates the sum of the diagonal elements of ${\bf{A}}$;
 $\mathcal{CN} (x;m,\sigma^2)$ denotes the complex Gaussian distribution with expectation $m$ and covariance $\sigma^2$.
 ${\bf D}_{N}$ denotes the $N\! \times\! N$ discrete Fourier transform matrix with $(m,n)$th element equal to $e^{-j2\pi(m\!-\!1)(n\!-\!1)/N}$.
  Finally, $\mathbb{E}(\cdot)$, ${\rm{var}}[\cdot]$, and $\Re\{\cdot\}$ denote the expectation, the variance, and the real part of the argument, respectively.

\vspace{-1mm}
\section{System Model}\label{S2}
\vspace{-1mm}

\begin{figure}[!tp]
\begin{center}
 \includegraphics[width=0.95\columnwidth,keepaspectratio]{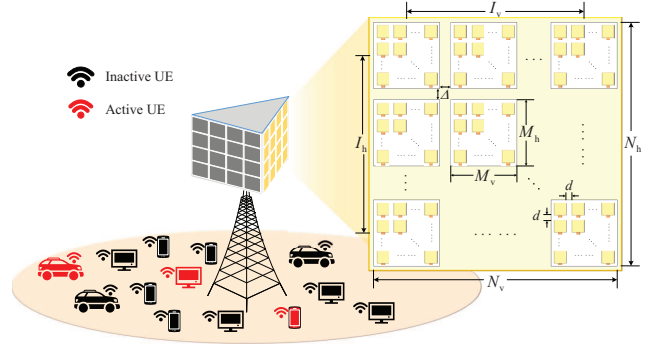}
\end{center}
 \captionsetup{font = {footnotesize}, singlelinecheck = off, justification = raggedright, name = {Fig.}, labelsep = period}
 \setlength{\abovecaptionskip}{-0.5mm}
\caption{Multi-panel mMIMO based massive IoT access system.}
 \label{FIG1}
\vspace{-5mm}
\end{figure}

 We consider a multi-panel mmWave/THz mMIMO system, where the BS equipped with a rectangular array serves $K$ potential single antenna UEs in uplink massive IoT access scenarios, as shown in Fig.~\ref{FIG1}.
 The BS adopts the multi-panel structure in conjunction with a partially-connected hybrid MIMO.
 The specific configuration of the rectangular antenna array is as follows.
 The number of subarray panels is $N_{\rm{P}}\! =\! I_{\rm h}I_{\rm v}$ with each of the subarray panels being a UPA, where $I_{\rm h}$ and $I_{\rm v}$ are the numbers of panels in the horizontal and vertical directions, respectively.
 We define $N_{\rm h}$ ($M_{\rm h}$) and $N_{\rm v}$ ($M_{\rm v}$) as the numbers of antennas in the horizontal and vertical directions of the rectangular array (subarray panel), respectively, i.e., $N_{\rm h}\! =\! I_{\rm h}M_{\rm h}$ and $N_{\rm v}\! =\! I_{\rm v}M_{\rm v}$.
 Therefore, the total number of antennas of the rectangular array is $N_{\rm BS}\! =\! N_{\rm h}N_{\rm v}$ ($M_{\rm BS}\! =\! M_{\rm h}M_{\rm v}$).
 The BS is equipped with $N_{\rm{P}}$ radio frequency (RF) chains, and each of them connects the corresponding subarray panel via the partially-connected phase shift network.
 Furthermore, the adjacent antenna spacing $d$ within each panel of Fig.~\ref{FIG1} is equal to $\lambda/2$, where $\lambda$ is the wavelength, and the adjacent panel spacing $\varDelta$ is equal to an integer multiple of $d$, yielding $\varDelta\! =\! Dd$ for $D\! \ge\! 2$.

 To combat the multipath effect at the BS caused by different scatterers in the communication environment, an orthogonal frequency-division multiplexing (OFDM) scheme having $N_{\rm c}$ subcarriers is applied for massive IoT access.
 Explicitly, $P$ subcarriers uniformly selected from the $N_{\rm c}$ available subcarriers can be utilized to transmit pilot signals for joint AUD and CE.
 Taking the special multi-panel mMIMO structure into consideration, the mmWave/THz channel ${\bf{h}}_{p,k}\! \in\! \mathbb{C}^{N_{\rm BS}}$ between the BS and the $k$th UE at the $p$th pilot subcarrier can be formulated as
\begin{equation}\label{channel_kp} 
 {\bf{h}}_{p,k} \!\!=\!\! \sum_{l=1}^{L} {\beta_{k,l} {\bf{a}}_{\rm MP}(\mu_{k,l},\nu_{k,l})
 e^{-\textsf{j}2 \pi \varpi_{k,l} \big( -{\textstyle{B_s \over 2}} +
 \big({\textstyle{pN_{\rm c} \over P}} - 1\big){\textstyle{B_s \over N_{\rm c}}} \big)}} ,
\end{equation}
 where $1\! \le\! k\! \le\! K$, $1\! \le\! p\! \le\! P$,
 $L$ is the total number of paths,
 ${\bf{a}}_{\rm MP}(\mu_{k,l},\nu_{k,l})\! \in\! \mathbb{C}^{N_{\rm BS}}$ is the array response vector evaluated at the horizontal and vertical virtual angles $\mu_{k,l}$ and $\nu_{k,l}$.
 Furthermore, $\beta_{k,l}\! \sim\! \mathcal{CN} (0, 1)$ and $\varpi_{k,l}$ denote the complex gain and path delay associated with the $l$th path, respectively,
 $B_{\rm{s}}$ is system bandwidth, and $N_{\rm c}/P$ is an integer.
 Specifically, by defining the horizontal and vertical virtual angles $\mu_{k,l}\! =\! \pi\sin\theta_{k,l}\cos\phi_{k,l}$ and $\nu_{k,l}\! =\! \pi\sin\phi_{k,l}$ with $\theta_{k,l}$ and $\phi_{k,l}$ being the azimuth and elevation angles, respectively,
 ${\bf{a}}_{\rm MP}(\mu_{k,l},\nu_{k,l})$ in (\ref{channel_kp}) can be acquired by the vectorization of ${\bf{A}}(\mu_{k,l},\nu_{k,l})\! =\! {\bf{a}}_{\rm h}(\mu_{k,l}){\bf{a}}_{\rm v}^{\rm T}(\nu_{k,l})$.
 Explicitly, we have ${\bf{a}}_{\rm{MP}}(\mu_{k,l},\nu_{k,l})\! =\! {\rm vec} \left[{\bf{A}}(\mu_{k,l},\nu_{k,l})\right]\! =\! {\bf{a}}_{\rm v}(\nu_{k,l})\! \otimes\! {\bf{a}}_{\rm h}(\mu_{k,l})$,
 while ${\bf{a}}_{\rm h}(\mu_{k,l})\! =\! {\bf{a}}_{\rm h}^{\rm I}(\mu_{k,l})\! \otimes\! {\bf{a}}_{\rm h}^{\rm M}(\mu_{k,l})\! \in\! \mathbb{C}^{N_{\rm h}}$ and ${\bf{a}}_{\rm v}(\nu_{k,l})\! =\! {\bf{a}}_{\rm v}^{\rm I}(\nu_{k,l})\! \otimes\! {\bf{a}}_{\rm v}^{\rm M}(\nu_{k,l})\! \in\! \mathbb{C}^{N_{\rm v}}$ are the horizontal and vertical steering vectors, respectively,
 in which the vectors ${\bf{a}}_{\rm h}^{\rm I}(\mu_{k,l})\! \in\! \mathbb{C}^{I_{\rm h}}$, ${\bf{a}}_{\rm h}^{\rm M}(\mu_{k,l})\! \in\! \mathbb{C}^{M_{\rm h}}$, ${\bf{a}}_{\rm v}^{\rm I}(\nu_{k,l})\! \in\! \mathbb{C}^{I_{\rm v}}$, and ${\bf{a}}_{\rm v}^{\rm M}(\nu_{k,l})\! \in\! \mathbb{C}^{M_{\rm v}}$ can be further written as
\begin{align} 
\vspace{-1mm}
 {\bf{a}}_{\rm h}^{\rm I}(\mu_{k,l}) =& \big[ 1,{e^{\textsf{j}(M_{\rm h}+D-1)\mu_{k,l}}}, \cdots , {e^{\textsf{j}(I_{\rm h}-1)(M_{\rm h}+D-1)\mu_{k,l}}} \big]^{\rm T} , \nonumber\\
 {\bf{a}}_{\rm h}^{\rm M}(\mu_{k,l}) =& \big[ 1, {e^{\textsf{j}\mu_{k,l}}}, \cdots , {e^{\textsf{j}(M_{\rm h}-1)\mu_{k,l}}} \big]^{\rm T} , \nonumber\\
 {\bf{a}}_{\rm v}^{\rm I}(\nu_{k,l}) =& \big[ 1,{e^{\textsf{j}(M_{\rm v}+D-1)\nu_{k,l}}}, \cdots , {e^{\textsf{j}(I_{\rm v}-1)(M_{\rm v}+D-1)\nu_{k,l}}} \big]^{\rm T} , \nonumber\\
 {\bf{a}}_{\rm v}^{\rm M}(\nu_{k,l}) =& \big[ 1, {e^{\textsf{j}\nu_{k,l}}}, \cdots , {e^{\textsf{j}(M_{\rm v}-1)\nu_{k,l}}} \big]^{\rm T} . \nonumber
\vspace{-5mm}
\end{align}
\vspace{-4mm}

{\color{black}
 Due to the inherently sporadic traffic pattern of typical massive IoT access, only a small fraction of the total UE population $K$ is activated, where the number of active UEs is $K_{\rm a}$ (usually $K_{\rm a}\! \ll\! K$).
 We define a binary activity indicator flag $\alpha_k$ as the activity of the $k$th UE, i.e., $\alpha_k\! =\! 1$ when the $k$th UE is active, and $\alpha_k\! =\! 0$ otherwise.
 The signal vector ${\bf{y}}_p^{(g)}\! \in\! \mathbb{C}^{N_{\rm P}}$ received at the BS from the $K$ UEs at the $p$th pilot subcarrier of the $g$th OFDM symbol can be expressed as
\begin{align}\label{y_pg_1} 
\vspace{-1mm}
 {\bf{y}}_p^{(g)} &= \big({\bf{W}}_{\rm RF}^{(g)} {\bf{W}}_{\rm BB}\big)^{\rm H} \sum\nolimits_{k=1}^K { \alpha_k {\bf{h}}_{p,k} s_{p,k}^{(g)} } + {\bf{n}}_p^{(g)} \nonumber\\
 &= \big({\bf{W}}_{\rm RF}^{(g)} {\bf{W}}_{\rm BB}\big)^{\rm H} {\bf{H}}_{p} {\bf{s}}_p^{(g)} + {\bf{n}}_p^{(g)},
\end{align}
 where ${\bf{W}}_{\rm RF}^{(g)}\! \in\! \mathbb{C}^{N_{\rm BS}\! \times\! N_{\rm P}}$ and ${\bf{W}}_{\rm BB}\! \in\! \mathbb{C}^{N_{\rm P}\! \times\! N_{\rm P}}$ denote the analog and digital combining matrices, respectively,
 ${\bf{H}}_p\! =\! \left[ \alpha_1 {\bf{h}}_{p,1}, \alpha_2 {\bf{h}}_{p,2},\! \cdots\! , \alpha_K {\bf{h}}_{p,K} \right]\! \in\! \mathbb{C}^{N_{\rm BS}\! \times\! K}$ is the channel matrix,
 ${\bf{s}}_p^{(g)}\! =\! \big[ s_{p,1}^{(g)}, s_{p,2}^{(g)},\! \cdots\! , s_{p,K}^{(g)} \big]^{\rm T}\! \in\! \mathbb{C}^{K}$ denotes the pilot signal vector, which is randomly selected from the columns of ${\bf D}_{K}$.
 and ${\bf{n}}_p^{(g)}\! =\! \big({\bf{W}}_{\rm RF}^{(g)} {\bf{W}}_{\rm BB}\big)^{\rm H} {\bf{\bar n}}_p^{(g)}$ is the noise vector with ${\bf{\bar n}}_p^{(g)}\! \in\! \mathbb{C}^{N_{\rm BS}}$ being the additive white Gaussian noise (AWGN), i.e., ${\bf{\bar n}}_p^{(g)}\! \sim\! \mathcal{CN} ({\bf{0}}_{N_{\rm BS}}, \sigma^2{\bf{I}}_{N_{\rm BS}})$.
 Observe that when $\alpha_{k}\! =\! 1$, the elements of the $k$th column of ${\bf{H}}_{p}$ are nonzero.
 With the definition of the binary activity indicator flag $\alpha_k$ and the combination between $\alpha_k$ and ${\bf h}_{p,k}$ in ${\bf{H}}_{p}$, the activity of UEs can be fully embedded in the channel matrix ${\bf{H}}_{p}$, which inspires us to jointly estimate the channel and detect the UEs' activity simultaneously.

 We assume the digital combining matrix to be an identity matrix, i.e., ${\bf{W}}_{\rm BB}\! =\! {\bf{I}}_{N_{\rm P}}$.
 To design ${\bf{W}}_{\rm RF}^{(g)}$, we first construct a partial unitary matrix ${\bf{Z}}^{(g)} \!=\! {\bf D}_{N_{\rm v}}\! \otimes \!{\bf D}_{N_{\rm h}} {\bf P}\! \in\! \mathbb{C}^{N_{\rm BS}\! \times\! N_{\rm P}}$, where the modulus of the elements in ${\bf{Z}}^{(g)}$ is 1 and ${\bf{P}}$ is a permutation matrix which consists of $N_{\rm P}$ columns randomly extracted from ${\bf{I}}_{N_{\rm BS}}$.
 For our partially-connected multi-panel array architecture at the BS,
 we initialize the $n_p$th column of ${\bf{W}}_{\rm RF}^{(g)}$ that corresponds to the $n_p$th RF chain as ${\bf{w}}_{n_p}^{(g)}\! =\! {\bf{0}}_{N_{\rm BS}}$,
 then let $[{\bf{w}}_{n_p}^{(g)}]_{{\cal I}_{n_p}}\! =\! {\textstyle{1 \over \sqrt{M_{\rm BS}}}}[{\bf{z}}_{n_p}^{(g)}]_{{\cal I}_{n_p}}$,
 where the ordered set ${\cal I}_{n_p}$ having a cardinality of $M_{\rm BS}$ denotes the antenna index of the $n_p$th subarray panel.
 Note that the design of fully-digital MIMO architecture does not have the constraints of ${\bf{W}}_{\rm RF}^{(g)}$.
 By contrast, this paper considers the multi-panel mMIMO with partially-connected hybrid MIMO architecture, which leads to the extra hardware constrains and poses the challenging on algorithm design.
 In Section III, we will formulate the joint AUD and CE scheme in the massive IoT access with multi-panel mMIMO system.

 }

\section{Proposed Joint AUD and CE Scheme}\label{S3}
{\color{black}
 In this section, we will formulate the joint AUD and CE scheme as a CS-based MMV problem with the utilization of the structured sparsity of massive IoT access channels.
 Furthermore, to solve this MMV-CS problem, the OAMP-EM-MMV algorithm is conceived where the EM algorithm learns the unknown parameters, i.e., the noise variance and the sparsity ratio.}
\vspace{-3mm}
\subsection{Formulation of  Massive IoT Access in Multi-Panel mMIMO}\label{S3.1}

{\color{black}
 We firstly focus on the received signal vector ${\bf{y}}_p^{(g)}$ in (\ref{y_pg_1}).
 By applying the vectorization rule ${\rm vec}({\bf A}{\bf B}{\bf C})\! = \! ({\bf C}^{\rm T} \! \otimes \! {\bf A})\! \cdot \!{\rm vec}({\bf B})$, the signal vector ${\bf{y}}_p^{(g)}$ can be rewritten as
}
\begin{equation}\label{y_pg_2} 
\vspace{-1mm}
 {\bf{y}}_p^{(g)} = {\bf{F}}_p^{(g)} {\bf{h}}_{p} + {\bf{n}}_p^{(g)} ,
\end{equation}
 where ${\bf{F}}_p^{(g)}\! =\! ({\bf{s}}_p^{(g)})^{\rm T}\! \otimes\! \big({\bf{W}}_{\rm RF}^{(g)}\big)^{\rm H}\! \in\! \mathbb{C}^{N_{\rm P}\! \times\! J}$,
 ${\bf{h}}_{p}\! =\! {\rm vec}({\bf{H}}_{p})\! \in\! \mathbb{C}^{J}$,
 and $J\! =\! K\!N_{\rm BS}$.
 The structured sparsity of the $p$th subchannel ${\bf{H}}_{p}$ is preserved in the vector ${\bf{h}}_{p}$.
{\color{black}
 Note that when the $k$th UE is active, the elements in ${\bf{h}}_{p}$ having indices from the $((k-1)N_{\rm BS}\!+\!1)$th to the $kN_{\rm{BS}}$th are nonzero, which inspires us that UEs' activity can be detected according to the position of non-zero elements and the structured sparsity of channel.}
 Furthermore, we consider the same signal vector used at all pilot subcarriers, i.e., ${\bf{s}}_p^{(g)}\! =\! {\bf{s}}^{(g)}$ and thus ${\bf{F}}_p^{(g)}\! =\! {\bf{F}}^{(g)}$ for $1\! \le\! p\! \le\! P$.
 By aggregating the received signals at the $P$ pilot subcarriers of the $g$th OFDM symbol as ${\bf{Y}}^{(g)}\! \in\! \mathbb{C}^{N_{\rm P}\! \times\! P}$, we have
\begin{equation}\label{Y_g} 
\vspace{-1mm}
 {\bf{Y}}^{(g)} = \left[ {\bf{y}}_1^{(g)}, {\bf{y}}_2^{(g)}, \cdots , {\bf{y}}_P^{(g)} \right]
 = {\bf{F}}^{(g)} {\bf{H}} + {\bf{N}}^{(g)} ,
\vspace{-1mm}
\end{equation}
 where ${\bf{H}}\! =\! \left[ {\bf{h}}_{1}, {\bf{h}}_{2},\! \cdots\! , {\bf{h}}_{P} \right]\! \in\! \mathbb{C}^{J\! \times\! P}$ and ${\bf{N}}^{(g)}$ denote the aggregated channel and noise matrices, respectively.

 It can be observed from (\ref{Y_g}) that, according to the identical UE activity $\alpha_k$, for $1\! \le\! k\! \le\! K$, observed at all subchannels, the aggregated channel matrix ${\bf{H}}$ exhibits the intrinsically structured sparsity.
 More explicitly, its columns, i.e., $\{{\bf{h}}_{p}\}_{p=1}^{P}$, have a common sparsity pattern (a. k. a. sparse support set) in the frequency domain, given by
\vspace{-3mm}
\begin{equation}\label{supp} 
\vspace{-1mm}
 {\rm supp} \{ {\bf{h}}_{1} \} = {\rm supp}\{ {\bf{h}}_{2} \} = \cdots = {\rm supp}\{ {\bf{h}}_{P} \},
\vspace{-1mm}
\end{equation}
 where ${\rm supp}\{\cdot\}$ denotes an ordered set consisting of the non-zero elements of the argument.
 {\color{black} Note that the support of ${\bf h}_{p}$ does not vary with the index of different subcarriers $p$, which can facilitate better CE performance.}

 Due to the limited observations in multi-panel mMIMO system relying on a partially-connected structure, we stack the received signal matrices in $G$ OFDM symbols, i.e., ${\bf{Y}}^{(g)}$ for $1\! \le\! g\! \le\! G$, to improve the joint AUD and CE performance.
 The stacked signal matrix ${\bf{Y}}\! \in\! \mathbb{C}^{Q\! \times\! P}$ can be expressed as
\begin{equation}\label{Y} 
\vspace{-1mm}
 {\bf{Y}} = \big[ ({\bf{Y}}^{(1)})^{\rm T}, ({\bf{Y}}^{(2)})^{\rm T}, \cdots , ({\bf{Y}}^{(G)})^{\rm T} \big]^{\rm T} = {\bf{F}} {\bf{H}} + {\bf{N}} ,
\vspace{-1mm}
\end{equation}
 where $Q\! =\! GN_{\rm P}$, while ${\bf{F}}\! =\! \big[ ({\bf{F}}^{(1)})^{\rm T}, \cdots , ({\bf{F}}^{(G)})^{\rm T} \big]^{\rm T}\!$ $\in\! \mathbb{C}^{Q\! \times\! J}$
 and ${\bf{N}}$ represent the sensing matrix and the stacked noise matrix, respectively.
 The sensing matrix ${\bf{F}}$ is a partial unitary matrix, which prompts us to design our solution developed from OAMP algorithm \cite{MaJ_Access17}.
 Since ${\bf{H}}$ exhibits the structured sparsity, the joint AUD and CE based on (\ref{Y}) is an MMV-CS problem associated with $Q\! \ll\! J$, which can be solved by the proposed OAMP-EM-MMV algorithm introduced in the next subsection.
 {\color{black}
 With the estimated channel $\widehat{\bf{H}}$, the support of $\widehat{\bf{H}}$ can be utilized to detect the activity of UEs, so the proposed solution is termed as a joint AUD and CE scheme.
 }

\vspace{-3mm}
\subsection{Proposed OAMP-EM-MMV Algorithm}\label{S3.2}
 The OAMP algorithm is developed from the AMP algorithm for solving the considered sparse signal recovery problem, while imposing a relaxed requirement on the sensing matrices \cite{MaJ_Access17}.
 When the sensing matrices are ill-conditioned transform matrices or partial unitary matrices, the performance of the AMP algorithm is not guaranteed, while the OAMP algorithm has improved robustness and performs still well as demonstrated in \cite{MaJ_Access17}.
 Specifically, the OAMP algorithm includes both a linear estimation (LE) module and a non-linear estimation (NLE) module, which are activated iteratively.
 The output of the NLE module is the MMSE estimate.
 Next, we elaborate on the proposed OAMP-EM-MMV algorithm.

 {\color{black}For the sparse channel matrix ${\bf{H}}$ in (4), the entries $h_{j,p}$ can be reasonably assumed to follow the Bernoulli-Gaussian distribution \cite{MaJ_Access17},
 and $\lambda_{j,p}$ denotes the sparsity ratio representing the non-zero probability of $h_{j,p}$.
 The proposed OAMP-EM-MMV algorithm involves $T$ iterations between the LE and NLE modules, and we focus our attention on the $t$th iteration.
 The linear MMSE (LMMSE) estimator and the mean error variance estimator of the LE module are listed  in the $5$th and $6$th lines of \textbf{Algorithm~\ref{Alg1}}, respectively.}

 {\color{black}The NLE module assumes that ${\bf{h}}_{p}$ is corrupted by an AWGN vector ${\bf{z}}_p$, i.e., we have ${\bf{r}}_{p}\! =\! {\bf{h}}_{p}\! +\! \tau_{p}{\bf{z}}_{p}$, where ${\bf{z}}_p\! \sim\! \mathcal{CN} ({\bf{0}}_{J}, {\bf{I}}_{J})$ is independent of ${\bf{h}}_{p}$.
 The mean error variance of the NLE module at the $t$th iteration $(v^2)^{t}$ can be further calculated as
\begin{equation}\label{v2}
\vspace{-1mm}
 (\upsilon_{p}^{2})^{t} = \big( {\textstyle{1 \over {\overline{\omega}}_{p}^{t}}} - {\textstyle{1 \over (\tau_{p}^{2})^{t}}} \big)^{-1},
\vspace{-1mm}
\end{equation}
 where ${\overline{\omega}}_{p}^{t}\! =\! {\textstyle{1 \over J}} \sum\nolimits_{j=1}^{J} {\rm{var}} \left[ h_{j,p} | r_{j,p}^{t} \right]$,
 and $r_{j,p}^{t}$ is the $j$th entry of ${\bf{r}}_{p}^{t}$.
 According to the {\emph{a priori}} distribution of $h_{j,p}$ and the NLE model, the {\emph{a posteriori}} distribution of $h_{j,p}$ can be represented as $p( h_{j,p} | r_{j,p} )\! =\! (1- \eta_{j,p}^{t}) \delta( h_{j,p} ) + {\eta_{j,p}^{t}} \mathcal{CN} (h_{j,p} ; 0 , (\psi^{2})^{t} ),$
 where $(\psi^{2})^{t}\! =\! {\textstyle{ \rho^2 (\upsilon_{p}^{2})^{t} \over \rho^2 + (\upsilon_{p}^{2})^{t} }}$, and
\begin{equation}\label{post_pi}
\vspace{-1mm}
\eta_{j,p}^{t} = b_{j,p}^{t} / ( a_{j,p}^{t} + b_{j,p}^{t}) ,
\vspace{-1mm}
\end{equation}
 with $a_{j,p}^{t}\!=\!{\textstyle{1- \lambda_{j,p} \over \pi ((\upsilon_{p}^{2})^{t}) }} e^{-{\textstyle{|r_{j,p}|^2 \over (\upsilon_{p}^{2})^{t} }}}$
 and $b_{j,p}^{t} \!=\! {\textstyle{\lambda_{j,p} \over \pi (\rho^2 + (\upsilon_{p}^{2})^{t}) }} e^{-{\textstyle{|r_{j,p}|^2 \over { \rho^2 + (\upsilon_{p}^{2})^{t} } }}}$.
 When $\eta_{j,p}^{t}$ tends to zero, $p ( h_{j,p} | r_{j,p} )$ can be approximately regarded as a Dirac function, and $h_{j,p}$ tends to zero.
 When $\eta_{j,p}^{t}$ tends to one, by contrast, $h_{j,p}$ tends to be nonzero.
 Therefore, $\eta_{j,p}^{t}$ is termed as the belief indicator (BI).
 The {\emph{a posteriori}} mean and variance can be expressed as
\vspace{-2mm}
\begin{align}
\vspace{-1mm}
\xi_{j,p}^{t}\! &=\! \mathbb{E} \left[ h_{j,p} | r_{j,p}^{t} \right]\! =\! \frac{b_{j,p}^{t}}{a_{j,p}^{t}+b_{j,p}^{t}} \kappa_{j,p}^{t} ,\label{post_E} \\
\vspace{-1mm}
\omega_{j,p}^{t}\! &=\! {\rm{var}} \left[ h_{j,p} | r_{j,p}^{t} \right] = \frac{b_{j,p}^{t} (\psi^{2})^{t}}{a_{j,p}^{t}+b_{j,p}^{t}} + \frac{a_{j,p}^{t} b_{j,p}^{t}\left| \kappa_{j,p}^{t} \right|^2}{(a_{j,p}^{t}+b_{j,p}^{t})^2} ,\label{post_var}
\end{align}
 where $\kappa_{j,p}^{t} = {\textstyle{\rho^2 \over \rho^2 + (\upsilon_{p}^{2})^{t} } r_{j,p}} $.}

\begin{algorithm}[t!]
\caption{OAMP-EM-MMV Algorithm}
\label{Alg1}
\begin{algorithmic}[1]
\REQUIRE~Received signal matrix ${\bf{Y}}$, sensing matrix ${\bf{F}}$, and maximum iterations $T$ \\
\ENSURE~Estimated channel $\widehat{\bf{H}}$, BIs $\eta_{j,p}, \forall j,p$
\STATE  ${\forall}j,p$: Calculate $\lambda_{j,p}^{0}$ in (\ref{ini_lambda}) and $({\sigma^2})^{0}$ in (\ref{ini_rho});\
\STATE  ${\forall}p$: Initialize ${\bf{r}}_{p}^{0}\! =\! {\bf{0}}_{J}$ and $(\upsilon_{p}^{2})^{0}\! =\! 1$;\
\FOR{$t\! =\! 1,\! \cdots\! ,T$}
\STATE \% LE module
{\color{black}
\STATE {\color{black} LMMSE: ${\forall}p$: ${\bf{r}}_{p}^{t}\! =\! {\bf{u}}_{p}^{t-1}\! +\! {\textstyle{J \over Q}} {\bf{F}}^{H} \left( {\bf{y}}_{p}\! -\! {\bf{F}} {\bf{u}}_{p}^{t-1} \right)$;}\
\STATE {\color{black} The mean error variance estimator: \\ ${\forall}p$: $(\tau_{p}^{2})^{t}\! =\! {\textstyle{ J-Q \over Q}} (\upsilon_{p}^{2})^{t-1}\! +\! {\textstyle{J \over Q}} (\sigma^2)^{t-1}$;}\
\STATE \% NLE module
\STATE ${\forall}j,p$: Calculate the {\emph{a posteriori}} mean $\xi_{j,p}^{t}$ in (\ref{post_E}) and variance $\omega_{j,p}^{t}$ in (\ref{post_var});\
\STATE ${\forall}p$: Calculate the mean error variance of the NLE $(\upsilon_{p}^{2})^{t}$ in (\ref{v2});\
\STATE ${\forall}j,p$: Update BI ${\eta}_{j,p}^{t}$ in (\ref{post_pi});\
}
\STATE \% EM module \\
\STATE ${\forall}j,p$: Update the parameters in (\ref{rho}) and (\ref{lambda_jp});\
\ENDFOR
\STATE ${\forall}j,p$: $\hat {h}_{j,p}\! =\! \xi_{j,p}^{T}$, and $\hat {h}_{j,p}$ is the $(j,p)$th element of $\widehat{\bf{H}}$.
\end{algorithmic}
\end{algorithm}

 As mentioned above, we revealed the theoretical basis process of the OAMP algorithm.
 The value of the noise variance $\sigma^2$ and the sparsity ratio $\lambda_{j,p}$ are required by the conventional OAMP algorithm.
 However, the exact values of these two parameters are difficult to obtain in practice, which motivates us to design adaptive parameter learning for enhancing the performance of the OAMP algorithm.
 Based on the above considerations, we integrate the EM algorithm into the OAMP algorithm.
 The EM algorithm is applied to estimate the unknown noise variance and sparsity ratio using the E step and M step, respectively,
\begin{align}
\vspace{-1mm}
{Q} \left( {\bm{\theta}} , {\bm{\theta}}^{t} \right) &= \mathbb{E} \left[ {\rm{ln}} p( {\bf{H}} , {\bf{Y}} ) | {\bf{Y}} ; {\bm{\theta}}^{t}  \right] , \label{Estep}  \\
{\bm{\theta}}^{t+1} &= \arg \max \limits_{{\bm{\theta}}} {Q} \left( {\bm{\theta}} , {\bm{\theta}}^{t} \right), \label{Mstep}
\end{align}
 where ${\rm{\mathbb{E}}} \left[ ( \cdot ) | {\bf{Y}} ;{\bm{\theta}}^{t}  \right] $ denotes the expectation conditioned on ${\bf{Y}}$ in conjunction with the parameters ${\bm{\theta}}^{t}\! = \! \{({\sigma^2})^{t},\lambda_{j,p}^{t}, {\forall}j,p\} $.
 The exact {\emph{a posteriori}} distribution required in (\ref{Estep}) is intractable, but we can approximate it from the OAMP algorithm.
 However, due to the multiple elements contained in ${\bm{\theta}}^{t}$ of (\ref{Mstep}), its joint optimization with ${\bm{\theta}}$ is difficult.
 Therefore, we adopt the so-called incremental EM algorithm, which estimates only a single parameter at each iteration, while keeping the others fixed.
 By taking the partial derivative of (\ref{Estep}) with respect to each element of ${\bm{\theta}}$ and setting the derivatives to zero,
 we obtain the update rules of ${\bm{\theta}}$ as
\begin{align}
\lambda_{j,p}^{t} &= \eta_{j,p}^{t-1}, {\forall}j,p , \label{lambda} \\
({\sigma^2})^{t} &= {\textstyle{1 \over P}}\{ \sum\nolimits_{p=1}^{P} {\textstyle{1 \over J}} \{ \sum\nolimits_{j=1}^{J} |r_{j,p}\! -\! \sum\nolimits_{p=1}^P f_{q,j} \xi_{j,p}^{t-1} |^2\! \} +\! {\overline{\omega}}_{p}^{t-1} \} , \label{rho}
\end{align}
 where $f_{q,j}$ is the $(q,j)$th element of ${\bf{F}}$.
 For the initialization of (\ref{lambda}) and (\ref{rho}) \cite{13TSP_EP}, the following expressions can be shown to be suitable
\begin{equation}\label{ini_lambda}
\lambda_{j,p}^{0}\! =\! {\textstyle{Q \over J}} \max \limits_{c>0} \frac{1\!-\!2 J [(1+c)^2 \Phi(-c)\!-\!c \phi(c) ]/Q}{1+c^2-2[(1\!+\!c)^2 \Phi(-c)\!-\!c \phi(c) ]}, \forall j,p, \\[-1mm]
\end{equation}
\begin{equation}\label{ini_rho}
({\sigma^2})^{0} = {\textstyle{1 \over P}} \sum\nolimits_{p=1}^{P} \frac{ || {\bf{y}}_{p} ||_{2}^{2} }{ \left( {\rm{SNR}}^{0} + 1 \right)Q },
\end{equation}
 where $\Phi (\cdot)$ and $\phi (\cdot)$ are the cumulative distribution function and probability distribution function of the standard normal distribution, respectively.
 Given that the initial signal-to-noise-ratio (i.e., ${\rm{SNR}}^{0}$) is usually unknown in practice, we set ${\rm{SNR}}^{0} = 100$, which is an appropriate empirical value.

 The OAMP algorithm assisted by the aforementioned EM algorithm is capable of solving the SMV problem.
 Furthermore, to solve the MMV problem in (\ref{Y}), the sparsity of ${\bf{H}}$ can be exploited and we adopt an innovative update rule to learn the structured sparsity.
 Since $\lambda_{j,p}$ represents the non-zero probability of $h_{j,p}$ and it is independently updated in (\ref{lambda_jp}), it is plausible that the sparsity of (\ref{supp}) cannot be exploited.
 In view of this fact, we can refine $\lambda_{j,p}$ as follows
\begin{equation}\label{lambda_jp}
{\lambda}_{j,1}^{t} = \cdots = {\lambda}_{j,P}^{t} = {\textstyle{1 \over P}} \sum\nolimits_{p=1}^{P} {\eta}_{j,p}^{t-1},
\end{equation}
 for exploiting the joint sparsity.
 {\color{black} Based on the aforementioned derivation and analysis, we summarize our OAMP-EM-MMV solution at a glance in \textbf{Algorithm~\ref{Alg1}}.


 After obtaining the CE result $\widehat{\bf{H}}$, we propose a pair of  AUD detectors based on $\widehat{\bf{H}}$ and ${\eta}_{j,p}$, respectively.
 Since the $P$ subchannels share the same support over all the $P$ subcarriers, we opt for the channel of arbitrary subcarrier, e.g., $p=1$, to detect the UEs' activity.
 Given the CE result $\widehat{\bf{H}}$, we may readily obtain the channel $\widehat{\bf{H}}_{1} \! \in\! \mathbb{C}^{N_{\rm BS}\! \times\! K}$, whose element is ${\hat{h}}_{n_{\rm{BS}},k}$.
 For the AUD, firstly a threshold function $r ( x; \epsilon) $ is defined beforehand, where $r ( x; \epsilon) $ equals 1 if $|x| > \epsilon$ and 0 otherwise.

 In accordance with the structured sparsity of the estimated channel matrix ${\widehat{\bf H}}$, we define the channel gain based activity detector (CG-AD) for AUD as follows
\begin{equation}
\widehat{\alpha}_{k} =
\left\{
\begin{aligned}
&1, {\textstyle{1 \over J}} \sum\nolimits_{n_{\rm{BS}}} \sum\nolimits_{k} r( {\hat{h}}_{n_{\rm{BS}},k} ; \epsilon_{\rm cg} ) \geq p_{\rm cg} ,     \\[-1mm]
&0, {\textstyle{1 \over J}} \sum\nolimits_{n_{\rm{BS}}} \sum\nolimits_{k} r( {\hat{h}}_{n_{\rm{BS}},k} ; \epsilon_{\rm cg} )   <  p_{\rm cg} ,
\end{aligned}
\right.
\vspace{-1mm}
\end{equation}
 where $\epsilon_{\rm cg} = 0.01 \max { \{ |{\hat{h}}_{j,k}|, \forall j,k \} }$ and $ p_{\rm cg} = 0.9$ \cite{KML_TSP20}.

 Furthermore, we define a BI based activity detector (BI-AD) as follows
\begin{equation}
\widehat{\alpha}_{k} =
\left\{
\begin{aligned}
&1, {\textstyle{1 \over J}} \sum\nolimits_{n_{\rm{BS}}} \sum\nolimits_{k} r( {\eta}_{n_{\rm{BS}},k} ; \epsilon_{\rm bi} ) \geq p_{\rm bi} , \\[-1mm]
&0, {\textstyle{1 \over J}} \sum\nolimits_{n_{\rm{BS}}} \sum\nolimits_{k} r( {\eta}_{n_{\rm{BS}},k} ; \epsilon_{\rm bi} ) < p_{\rm bi} ,
\end{aligned}
\right.
\vspace{-1mm}
\end{equation}
 where $\{{\eta}_{n_{\rm{BS}},k}, \forall n_{\rm{BS}},k\}$ can be obtained from ${\eta}_{1,1}$ to ${\eta}_{J,1}$.
 
 For our channel model, we set $\epsilon_{\rm bi}$ to $0.5$ for convenience{\footnote{\color{black}The choice of $\epsilon_{\rm bi}$ can be further optimized according to the cost of missed detection and false alarm required by the practical communication systems.}}.

\color{black}
\vspace{-3mm}
\section{Simulation Results}\label{S4}
\vspace{-1mm}
 In this section, we evaluate the performance of the proposed joint AUD and CE scheme based on multi-panel mMIMO aided massive IoT access.
 In our simulations, the carrier frequency, bandwidth, and the number of subcarriers are $30$\,GHz, $B_s\! =\! 1$\,GHz, and $N_{\rm{c}}\! =\! 256$, respectively.
 For the multi-panel mMIMO array at the BS, we use ${I}_{\rm{v}}\! =\! {I}_{\rm{h}}\! =\! 4$,
 that is ${N}_{\rm{P}}\! =\! 16$ panels, and $M_{\rm{h}}\! =\! M_{\rm{v}}\! =\! 2$ for each panel,
 so that the total number of antennas in this multi-panel mMIMO is $N_{\rm{BS}}\!=\! 64$.
 The adjacent panel spacing is $\varDelta\! =\! 6d$, i.e., $D\! =\!6$.
 Furthermore, in the channel, $L\! =\! 4$, and the path delay $\varpi_{k,l}$ follows the uniform distribution ${\cal U}[0,32/B_s]$.
 The maximum number of iterations in \textbf{Algorithm~\ref{Alg1}} is $T\! =\! 100$ and ${\rm{SNR}}\! =\! 30$\,dB.
 The AUD error probability and the CE MSE defined in \cite{KML_TSP20} are used as our performance metrics.
 Based on our simulation parameters, the transmission delay of an OFDM symbol is equal to $0.288$\,microsecond ($\mu {\rm{s}}$).

\begin{figure}[!tp]
\begin{center}
 \includegraphics[width=0.9\columnwidth,keepaspectratio]{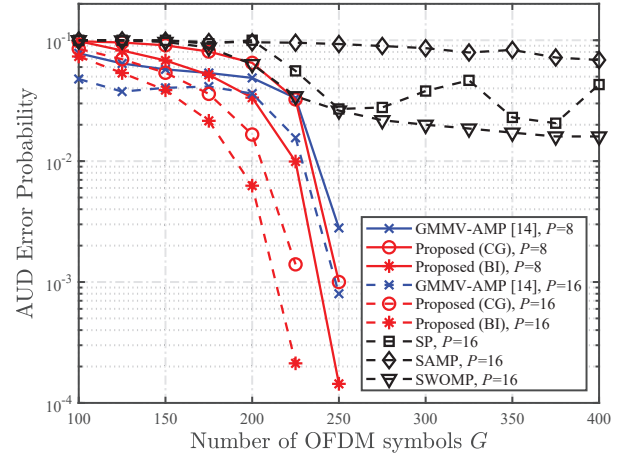}
\end{center}
 \captionsetup{font = {footnotesize}, singlelinecheck = off, justification = raggedright, name = {Fig.}, labelsep = period}
 \setlength{\abovecaptionskip}{-0.5mm}
\caption{{\color{black}AUD performance comparison of different schemes versus $G$, where $K\!=\!500$ and $K_a\! =\! 50$, and we consider the cases of $P\!=\!8$ and $P\!=\!16$.}}
 \label{FIG2}
\vspace{-5mm}
\end{figure}

\begin{figure}[!tp]
\begin{center}
 \includegraphics[width=0.9\columnwidth,keepaspectratio]{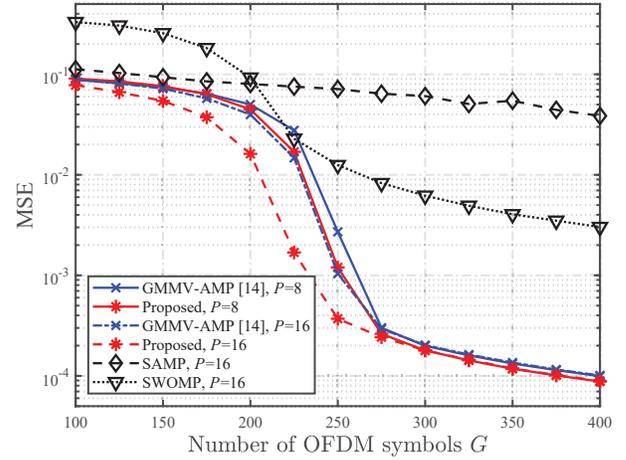}
\end{center}
 \captionsetup{font = {footnotesize}, singlelinecheck = off, justification = raggedright, name = {Fig.}, labelsep = period}
 \setlength{\abovecaptionskip}{-0.5mm}
\caption{{\color{black}CE performance comparison of different schemes versus $G$, where $K\!=\!500$ and $K_a\! =\! 50$, and we consider the cases of $P\!=\!8$ and $P\!=\!16$.}}
 \label{FIG3}
\vspace{-5mm}
\end{figure}

 Fig.~\ref{FIG2} compares the AUD performance of different schemes versus the number of OFDM symbols $G$.
 {\color{black}
 In Fig.~\ref{FIG2} and Fig.~\ref{FIG3}, we set $K\!=\!500$ and $K_a\! =\! 50$, and we consider the cases of $P\! =\! 8$ and $P\! =\! 16$.}
 We observe from Fig.~\ref{FIG2} that the proposed OAMP-EM-MMV algorithm outperforms the other three greedy algorithms (namely the SAMP, SP, and SWOMP algorithms utilized in \cite{KML_TSP20} as baseline schemes), despite using less pilot subcarriers, and has a significant advantage over the GMMV-AMP algorithm \cite{KML_TSP20}.
 Furthermore, for the proposed OAMP-EM-MMV algorithms relying on the CG-AD and BI-AD, the AUD performance of BI-AD is distinctly better than that of CG-AD.
 When $G\! \geq\! 225$, the AUD performance of the CG-AD and BI-AD for $P\! =\! 16$ tends to zero quite rapidly.
 In the case of $P\! =\! 16$, the AUD performance of the proposed BI-AD tends to zero rapidly when $G\! \geq\! 250$.
 Hence, all the UEs can be detected correctly within the access latency of $72\,\mu{\rm s}$.

 Fig.~\ref{FIG3} compares the MSE performance of the CE versus the number of OFDM symbols $G$.
 In Fig.~\ref{FIG3}, the MSE performance of the proposed OAMP-EM-MMV algorithm is seen to be superior to the other baseline algorithms, especially when $P\! =\! 16$.
 The CE accuracy of the proposed algorithm relying on less pilot subcarriers, i.e., $P\! =\! 8$, will be better than that of the baseline algorithms using $P\! =\! 16$.
 When $200\! \le\! G\! \le\! 275$, observe from Fig.~\ref{FIG3} that the MSE curves of the algorithms based on the message passing method decays rapidly,
 while these MSE curves will almost overlap when $G$ is large enough (e.g., $G\! >\! 275$).
 It becomes clear from Fig.~\ref{FIG2} and Fig.~\ref{FIG3} that the access latency to achieve reliable joint AUD and CE performance is less than $79.2 \mu {\rm{s}}$, which can meet the latency requirements of the IoV.

\begin{figure}[!tp]
\begin{center}
 \includegraphics[width=0.9\columnwidth,keepaspectratio]{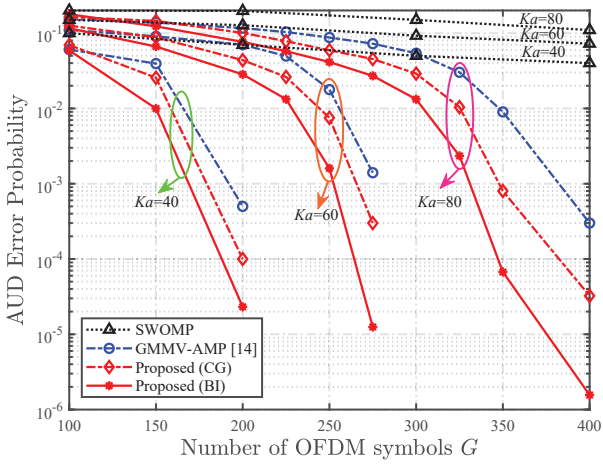}
\end{center}
 \captionsetup{font = {footnotesize}, singlelinecheck = off, justification = raggedright, name = {Fig.}, labelsep = period}
 \setlength{\abovecaptionskip}{-0.5mm}
\caption{{\color{black}AUD performance comparison of different schemes versus $G$, where $P\!=\!16$ and $K\!=\!400$, and we consider the cases of $K_a\!=\!40$, $60$, and $80$.}}
 \label{FIG4}
\vspace{-3mm}
\end{figure}

\begin{figure}[!tp]
\begin{center}
 \includegraphics[width=0.9\columnwidth,keepaspectratio]{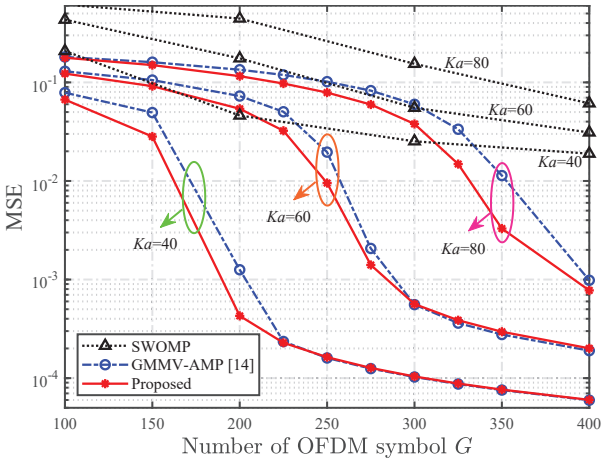}
\end{center}
 \captionsetup{font = {footnotesize}, singlelinecheck = off, justification = raggedright, name = {Fig.}, labelsep = period}
 \setlength{\abovecaptionskip}{-0.5mm}
\caption{{\color{black}CE performance comparison of different schemes versus $G$, where $P\!=\!16$ and $K\!=\!400$, and we consider the cases of $K_a\!=\!40$, $60$, and $80$.}}
 \label{FIG5}
\vspace{-3mm}
\end{figure}

{\color{black}
Fig.~\ref{FIG4} compares the AUD performance of different schemes versus the number of OFDM symbols $G$ with different ratios of active UEs.
In Fig.~\ref{FIG4} and Fig.~\ref{FIG5}, we consider $P\!=\!16$ and $K\!=\!400$,
and $K_a$ is set to 40, 60 and 80 so the sparsity ratio is 10\%, 15\%, and 20\%, respectively.
In the cases of $K_a\!=\!40$ and $K_a\! =\! 60$, the AUD error probability becomes very small when the number of OFDM symbols $G$ exceeds 150 and 275, respectively.
It can be observed from Fig.~\ref{FIG4} that the AUD performance of each algorithm deteriorates as the number of the active UEs and the sparsity ratio increase.
While given one specific value of $K_a$, the proposed OAMP-EM-MMV algorithm is obviously superior to other baseline algorithms, which demonstrates the robustness of the proposed OAMP-EM-MMV algorithm.
Furthermore, for the proposed OAMP-EM-MMV algorithm relying on the CG-AD and BI-AD, the AUD performance of BI-AD is better than that of CG-AD in the case of different numbers of the active UEs, which indicates that EM algorithm can update the sparsity ratio robustly when the sparsity level changes.\\

Fig.~\ref{FIG5} compares the MSE performance of the CE versus the number of OFDM symbols $G$ with different numbers of the active UEs.
In the cases of $K_a\! =\! 40$ and $K_a\! =\! 60$, the MSE declines rapidly when $150 \leqslant \! G\! \leqslant \!200$ and $225\! \leqslant \! G \! \leqslant300$, respectively.
Given one specific value of $K_a$, the MSE performance of the proposed OAMP-EM-MMV algorithm is superior to other baseline algorithms in the cases of different numbers of the active UEs.
Furthermore, the simulation results of Fig.~\ref{FIG4} and Fig.~\ref{FIG5} demonstrate the superiority of the combination between the OAMP algorithm and the EM algorithm.
}

\vspace{-2mm}
\section{Conclusions}\label{S5}
\vspace{-1mm}
 In this paper, we have proposed a CS-based joint AUD and CE scheme for massive IoT access relying on mmWave/THz multi-panel mMIMO.
 Since the multi-panel mMIMO is a kind of partially-connected hybrid MIMO, the existing AUD and CE schemes designed for fully-digital MIMO can not perform well.
 Specifically, by designing the uplink combining matrix and exploiting the structured sparsity of the uplink massive IoT access channels, the joint AUD and CE problem can be formulated as an MMV-CS problem.
 We further develop an OAMP-EM-MMV algorithm to solve this problem by utilizing the EM algorithm to learn the {\emph{a priori}} parameters, i.e., the noise variance and the sparsity ratio.
 Our simulation results have demonstrated that the proposed OAMP-EM-MMV algorithm based joint AUD and CE scheme achieves better AUD and CE performance than the state-of-the-art schemes.

\ifCLASSOPTIONcaptionsoff
  \newpage
\fi

\end{document}